%% file: main.tex
\documentclass[runningheads]{llncs}
\usepackage[T1]{fontenc}

\newif\ifanonymous
\anonymousfalse

\usepackage{cite}
\usepackage{graphicx}
\usepackage{cleveref}
\usepackage{tabularx}
\usepackage{booktabs}
\usepackage{array}
\usepackage{makecell}
\usepackage{xcolor}
\newcommand{\rev}[1]{\textcolor{black}{#1}}
\begin{document}
\title{Through Human Eyes and Machine Eyes: Understanding View Mismatch in Video See-Through Extended Reality}

\titlerunning{View Mismatch in Video See-Through XR}


\ifanonymous
\author{Anonymous Author(s)}
\authorrunning{Anonymous}
\institute{Anonymous Institution}
\else
\author{Yanming Xiu\inst{}}
\authorrunning{Y. Xiu}
\institute{Department of Electrical and Computer Engineering, Duke University \\ Durham, NC, 27707, USA}
\fi

\maketitle              
\begin{abstract}


Video see-through extended reality (VST XR) systems commonly use headset screenshots or captured frames as proxies for the user's first-person visual context. However, the system-captured view and the user's effective visible field do not necessarily coincide: a screenshot records a rectangular machine-readable frame, whereas the user's effective visible region can be more constrained and non-rectangular. This paper studies this human-system view mismatch in VST XR. We formalize the relationship between the system-captured region and the human-visible region by defining their co-visible, system-only, and human-only regions. \rev{We then conduct a pilot-level boundary measurement on Meta Quest 3, revealing a clear mismatch between the rectangular screenshot frame and the approximate human-visible boundary. Building on this model, we analyze how view mismatch can affect screenshot-based XR sensing and downstream vision-language model tasks. Through four representative case studies, we illustrate potential risks and failure modes including prompt injection, privacy leakage, human-invisible information bias, and missing human-visible information. Our results show that view mismatch is not only a geometric artifact, but can also introduce security, privacy, and reliability concerns for AI-integrated VST XR systems.}

\keywords{Extended Reality  \and Video See-through \and Egocentric Vision \and Human Field of View \and XR Security \and XR Privacy \and User Experience}
\end{abstract}


\input{sec/1_intro}
\input{sec/2_related}
\input{sec/3_definition}
\input{sec/5_threat}
\input{sec/6_implement}

\input{sec/7_discussion}
\input{sec/8_conclusion}



%
%
%
\bibliographystyle{splncs04}

\input{reference.bbl}
\end{document}

%% file: sec/1_intro.tex

\section{Introduction}

Video see-through (VST) XR systems, such as Meta Quest and Apple Vision Pro, reconstruct the physical world through cameras and composite virtual content into the rendered view~\cite{de2024visual, westermeier2024assessing}. Compared with optical see-through displays, VST systems provide greater control over the final visual output and virtual-content rendering~\cite{rolland2000optical, itoh2021towards, macedo2021occlusion, ishihara2023integrating}. As a result, system-rendered and system-captured views are often treated as proxies for what the user is assumed to see.

\begin{figure}[t]
\includegraphics[width=0.9\linewidth]{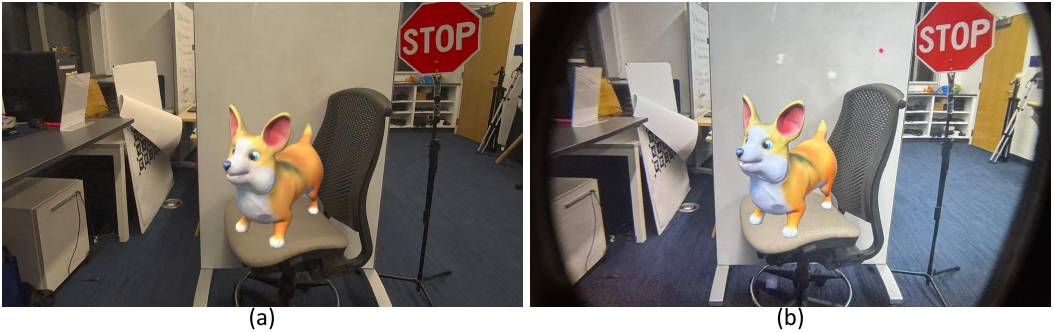}
\centering
\vspace{-0.5cm}
\caption{An illustrative example of human-system view mismatch in VST XR. Both images are captured from the left-eye perspective. (a) The system-captured view from Meta Quest 3. (b) An approximate human-visible view captured by placing a USB camera behind the headset lens.}
\label{fig:view_mismatch_example}
\vspace{-0.6cm}
\end{figure}

Many XR systems use head-mounted displays' built-in camera or screenshot functionality to obtain a user's first-person view~\cite{chen_2026_CVPR, gonzalez2020enhanced, pigny2019using, lee2025sensible, yu2026reality}. These captured views support downstream tasks such as scene understanding, safety analysis, and human-AI interaction, and content appearing in the captured frame are implicitly assumed to also be visible to the user. However, this assumption overlooks a fundamental mismatch between system-captured views and human-visible content in VST XR. The system records a rectangular rendered frame, whereas the user's effective visible field is constrained by the headset optics, lenses, and viewing conditions. As a result, peripheral or corner regions in the captured frame may remain outside the user's perceived field of view, creating system-visible but human-invisible areas. Fig.~\ref{fig:view_mismatch_example} illustrates this mismatch using a Meta Quest 3 screenshot and an approximate human-visible view captured from behind the headset lens. Although this camera-based view is not an exact measurement of human vision, it provides an intuitive visualization of the mismatch.

Although this difference can be observed qualitatively, its magnitude and system-level consequences have not been systematically studied. In this paper, we investigate this mismatch in VST XR and examine its implications for XR evaluation, security, privacy, and reliability. Content in screenshot-visible but human-invisible regions may be processed by downstream systems without being perceived by the user, potentially leading to incorrect scene interpretation, privacy leakage, or adversarial effects such as hidden visual prompt injection. Conversely, human-visible content may be omitted from system-captured screenshots, making them incomplete records of the user's visual experience. \rev{To study these issues, we formalize the relationship between system-captured and human-visible regions, conduct a pilot measurement on Meta Quest 3, and analyze representative risks arising from this mismatch. Our contributions are as follows:}

\vspace{-0.2cm}

\begin{itemize}
\item We formally model the mismatch between system-captured screenshots and the human-visible regions in VST XR.
\item We conduct a pilot-level boundary measurement on Meta Quest 3 to provide preliminary empirical evidence of the mismatch and obtain a setup-specific estimate of the human-visible boundary.
\item We develop a threat and impact model for XR systems that rely on screenshots as proxies for human-visible content.
\item \rev{We implement and test representative cases, including prompt injection, privacy leakage, human-invisible information bias, and missing human-visible information, to illustrate potential consequences of both system-visible human-invisible content and human-visible system-invisible content.}
\end{itemize}


%% file: sec/2_related.tex
\section{Related Work}

\label{sec:related}

\vspace{-0.1cm}
\subsection{Human Visual Fields and Visibility in Head-Mounted Displays}
\vspace{-0.1cm}

Human visual fields have long been studied in clinical vision science and ophthalmology. Prior work on visual-field examination and perimetry shows that human visibility is asymmetric rather than rectangular, with different extents across directions and distinctions between central and peripheral vision~\cite{clark1990clinical,grzybowski2009harry}. This is directly relevant to XR, where the effective visible region need not match a rectangular image frame.

In HMDs, visibility is further shaped by display optics, eye position, eye relief, interpupillary distance, and device fit. Prior work has compared optical see-through and video see-through HMDs~\cite{rolland2000optical} and shown that HMD visibility depends on optical and display configuration rather than the rendered image alone~\cite{rolland2005head}. Recent studies of video see-through HMDs have also examined distance perception, visual quality, contrast and color perception, user satisfaction, and spatial interaction accuracy~\cite{de2024visual,pfeil2021distance,de2024evaluating,wang2026perceptual,vona2025comparing}. These studies establish that XR visibility and perception are shaped by human vision, HMD optics, and the camera-display pipeline. However, prior work has primarily focused on perceptual quality and interaction performance. In contrast, we study whether system-captured views faithfully represent the region effectively visible to the user when used in downstream analysis pipelines.

\vspace{-0.2cm}
\subsection{System-Captured Views in Video See-Through XR Analysis}
\vspace{-0.1cm}

Prior work in VST and augmented virtuality has long used camera-captured first-person views for automated visual analysis. Examples include user segmentation in video see-through augmented virtuality~\cite{pigny2019using}, egocentric arm segmentation~\cite{gonzalez2020enhanced}, and upper-limb segmentation from egocentric RGB views~\cite{gruosso2021exploring}. These studies show that system-captured first-person views were already used as machine-readable representations of the user's body and surrounding context before the recent adoption of large multimodal models.

Recent XR systems increasingly use  vision language models (VLMs) to interpret captured first-person views. Examples include VLM-based obstruction detection~\cite{xiu2025vision}, Sensible Agent and Reality Copilot for proactive assistance and human-AI collaboration~\cite{lee2025sensible,yu2026reality}, Frame-First AR for egocentric scene-graph construction~\cite{naghadeh2026frame}, and VLM-based analysis of AR-generated scenes~\cite{duan2025advancing,duan2025probing}. Together, these works show that system-captured first-person views increasingly serve as inputs to downstream XR perception and reasoning. However, this use can implicitly assume that the captured view represents the region effectively visible to the user. Our work examines this assumption by considering both directions of mismatch: captured views may contain human-invisible content or omit content that remains human-visible.

\vspace{-0.2cm}
\subsection{Safety, Privacy, and Reliability Risks in AI-Integrated XR}
\vspace{-0.1cm}

Prior work has identified several safety, privacy, and reliability risks in AI-integrated XR. Prompt-injection attacks can manipulate multimodal models through adversarial text-bearing objects in physical or 3D environments~\cite{li2026extended}. Privacy-aware systems such as PRISM-XR preprocess headset-captured visual inputs before cloud transmission~\cite{chen2026prism}, while semantic context-aware methods detect privacy-sensitive visual content in AR~\cite{liu2026see}. Other work studies task-detrimental or misleading AR content, including obstruction and visual information manipulation attacks~\cite{xiu2025vision,xiu2025detecting,chen_2026_CVPR}.

These studies demonstrate risks arising from malicious prompts, privacy-sensitive content, and misleading AR scenes. However, they generally analyze the content present in captured views without explicitly considering whether those views match the effective human-visible region. Our work examines this additional source of risk, where system-visible but human-invisible content can influence downstream models, while human-visible but system-invisible content can make screenshot-based analysis incomplete.

%% file: sec/3_definition.tex
\vspace{-0.2cm}
\section{View Mismatch: Definition and Measurement}
\vspace{-0.2cm}

\label{sec:definition}

In this section, we first define view mismatch between the system-captured screenshot region and the human-visible region. We then present a pilot-level boundary measurement on Meta Quest 3 to examine whether the screenshot boundary aligns with an approximate human-visible boundary. This measurement is intended as an initial empirical probe rather than an exhaustive characterization of user- and device-dependent factors.

\vspace{-0.2cm}
\subsection{Problem Definition}
\vspace{-0.1cm}

We model view mismatch in video see-through XR as a discrepancy between the visual information captured by the system and the visual information effectively visible to the user. Let $I$ denote the visual information space of a video see-through scene at a given time. We define the system-captured region as $S \subseteq I$ and the human-visible region as $H \subseteq I$. Here, $S$ represents information available through screenshots, exported frames, or system-accessible rendered views, while $H$ represents information visible to the user through the headset optics, display, lenses, eye position, and device fit.

Under the common proxy assumption, screenshots are treated as approximate representations of the user's visual field, i.e., $S \approx H$. However, this assumption does not always hold in video see-through XR. \rev{Although developers may crop captured frames or choose different aspect ratios, such changes only alter the geometry of the system-captured region $S$ and do not by themselves guarantee alignment with the human-visible region $H$.} The shared region is defined as $R_c = S \cap H$, which contains visual information that is both system-visible and human-visible. The two non-overlapping regions are $R_s = S \setminus H$ and $R_h = H \setminus S$.

\begin{figure}[t]
\includegraphics[width=0.7\linewidth]{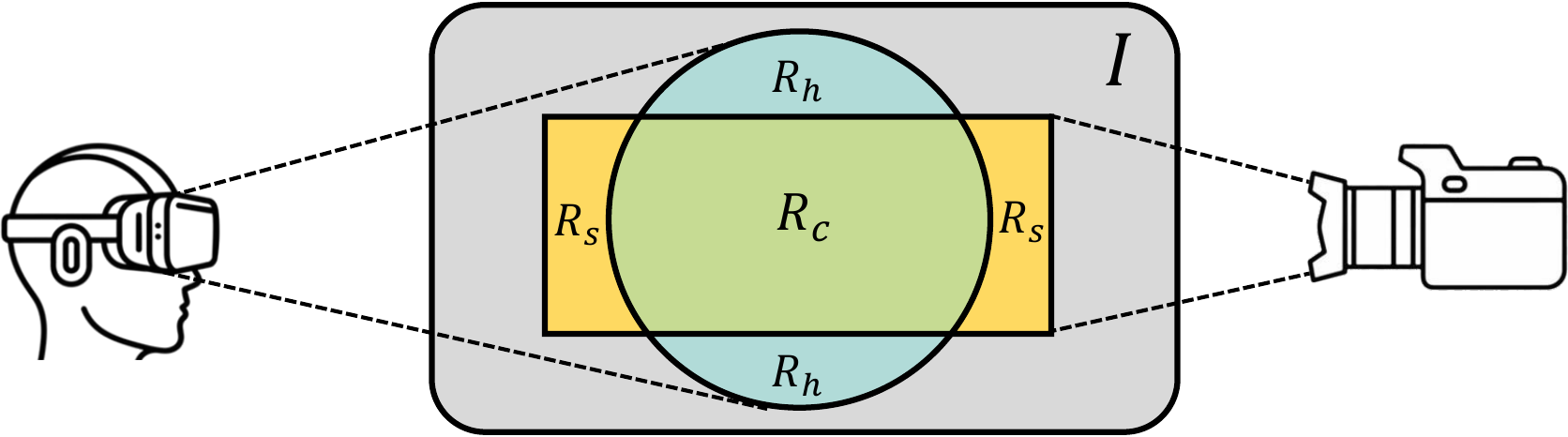}
\centering
\vspace{-0.2cm}
\caption{Conceptual illustration of view mismatch in video see-through XR. The yellow rectangle denotes the system-captured region $S$, while the blue area denotes the human-visible region $H$. Their intersection in green forms the co-visible region $R_c$. The system-only region is $R_s = S \setminus H$ and the human-only region is $R_h = H \setminus S$. }
\label{fig:definition_venn}
\vspace{-0.7cm}
\end{figure}

Here, $R_s$ denotes the system-visible but human-invisible region, where visual content can be captured and processed by the system while remaining outside the user's effective field of view. In contrast, $R_h$ denotes the human-visible but system-invisible region, where content perceived by the user may be missing from screenshots, logs, or downstream analysis. Fig.~\ref{fig:definition_venn} illustrates these three regions and their relationship to the system-captured and human-visible views.

Together, these regions partition the union of system-captured and human-visible information:

\vspace{-0.2cm}

$$
S \cup H = R_c \cup R_s \cup R_h.
$$

For an individual visual element $x \in S \cup H$, its visibility relation is determined by the region to which it belongs: $x \in R_c$ is visible to both the system and the user, $x \in R_s$ is visible only to the system, and $x \in R_h$ is visible only to the user. In the following sections, we examine how both $R_s$ and $R_h$ can affect screenshot-based XR sensing, downstream machine perception, and the reliability of human-visible context modeling.

\vspace{-0.2cm}
\subsection{Human-View Boundary Measurement Setup}
\vspace{-0.1cm}

Based on the definition of view mismatch between the system-captured and the human-visible region, we conducted a pilot-level boundary measurement to estimate how these two regions differ on a representative VST XR headset. We used Meta Quest 3 as the measurement device because it is a widely used consumer headset and provides a developer-accessible platform for mixed reality applications. Our goal was not to obtain a device-independent or user-independent optical model, but to obtain a setup-specific estimate of the human-visible boundary that can be compared with the system-captured screenshot frame.

\begin{figure}[t]
\includegraphics[width=0.88\linewidth]{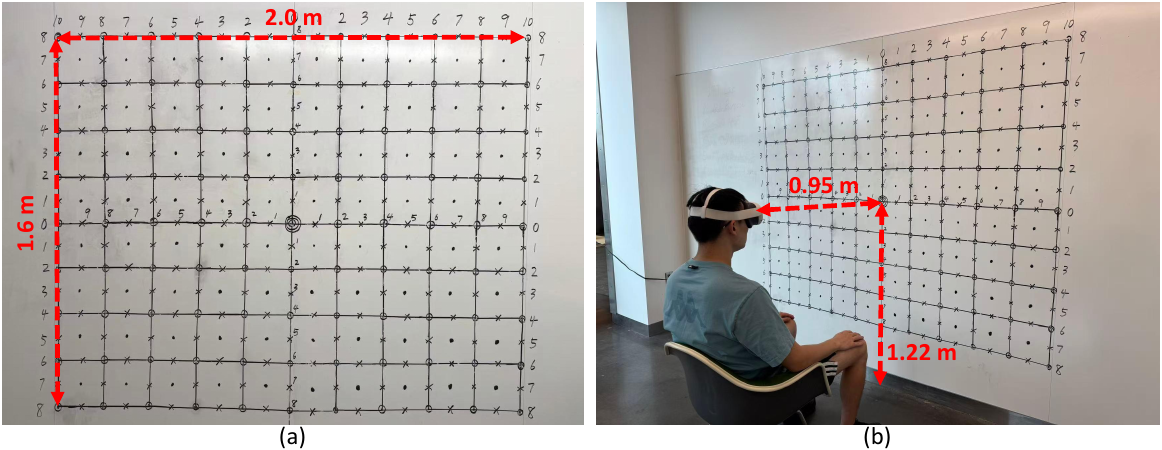}
\centering
\vspace{-0.4cm}
\caption{Whiteboard-based setup for pilot-level human-view boundary measurement.
(a) The manually drawn grid used as the spatial reference.
(b) The physical setup, where the observer wearing Meta Quest 3 sat on a fixed chair facing the grid.}
\label{fig:measurement_setup}
\vspace{-0.65cm}
\end{figure}

We designed a whiteboard-based calibration setup for this measurement. As shown in Fig.~\ref{fig:measurement_setup}(a), we manually drew a grid region on a whiteboard. The grid covered approximately 2.0 meters in width and 1.6 meters in height, with adjacent grid points spaced 10~cm apart and major grid lines drawn every 20~cm. The center of the grid was positioned 1.22 meters above the floor. During the measurement, the observer sat on a fixed chair facing the whiteboard, kept the head level, and adjusted the seated position so that the grid center was roughly aligned with the center of the headset view. The distance between the whiteboard and the front of the headset's main camera was 0.95 meters. This grid-based setup is consistent with prior HMD calibration practice, where regular grid patterns or planar calibration targets are used as spatial references for estimating display-viewpoint relationships~\cite{hua2002calibration}. The overall physical setup is shown in Fig.~\ref{fig:measurement_setup}(b).

\rev{The headset was worn in a normal-use condition: it was adjusted so that it did not feel loose or slide downward, while also avoiding excessive tightness or pressure on the head. This criterion was intended to approximate a typical comfortable wearing condition rather than an extreme fit. The observer was the author of this paper, whose interpupillary distance (IPD), as previously measured by a professional optometrist and reported by the observer, was 63~mm. The Meta Quest 3 IPD was therefore adjusted to 63~mm using the headset's physical IPD adjustment wheel. The human-visible boundary was measured under binocular viewing, consistent with normal headset use and everyday visual perception, whereas the system-captured screenshot was obtained from the Meta Quest 3 built-in screenshot function, which under the default capture configuration exports the left-eye view at 3840$\times$2160 resolution with a 16:9 aspect ratio.}

\rev{The measurement procedure consisted of two steps. First, the observer used the built-in screenshot function of Meta Quest 3 to save the system-captured view of the whiteboard grid, which provides the system-captured region $S$ in the image coordinate system. Second, the observer maintained visual fixation on the center of the whiteboard grid while remaining seated in the fixed chair with the body facing forward and the head kept approximately steady. No mechanical head restraint was used, so small involuntary head movements may have occurred during the measurement. Under this condition, a grid point was classified as visible if the observer could still perceive it while maintaining fixation on the grid center and keeping the body and head approximately fixed; otherwise, it was classified as invisible. The observer manually recorded the approximate boundary formed by the outermost grid points that remained visible, and these samples were used to estimate the human-visible region $H$ and compare it with the screenshot frame. Because the measurement relies on manually observed grid boundaries under a single binocular viewing condition and without mechanical head stabilization, we treat the resulting contour as an approximate pilot-level estimate rather than an exact measurement of the optical field of view.}

\vspace{-0.2cm}
\subsection{Measurement Results and Analysis}
\vspace{-0.1cm}

\begin{figure}[t]
    \centering
    \includegraphics[width=0.86\linewidth]{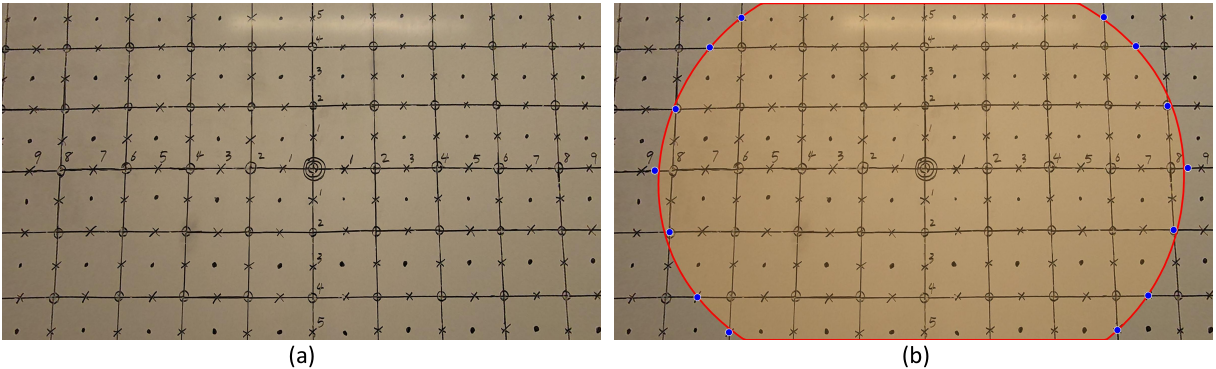}
    \vspace{-0.4cm}
    \caption{Pilot-level boundary measurement result. (a) The system-captured region obtained from the Meta Quest 3 screenshot. (b) The fitted approximate human-visible boundary overlaid on the screenshot using recorded boundary samples.}
    \label{fig:measurement_result}
    \vspace{-0.6cm}
\end{figure}

We conducted the pilot measurement using the setup described above. Fig.~\ref{fig:measurement_result}(a) shows the captured whiteboard grid, which serves as the system-captured region $S$ in our analysis. We then recorded the approximate boundary of the grid points that remained visible to the observer through the headset. In the whiteboard coordinate system, the observed boundary was approximately elliptical. On the left side of the visible boundary, the sampled boundary points were approximately $(-5, 6)$, $(-6, 5)$, $(-7, 4)$, $(-8, 2)$, $(-8.5, 0)$, $(-8, -2)$, $(-7, -4)$, $(-6, -5)$, and $(-5, -6)$. The right side of the boundary was approximately symmetric with respect to the vertical axis of the grid. These boundary samples were already close to the edge of the observer's visible field. When the observer attempted to inspect these points, the corresponding visual content appeared near the optical periphery and was accompanied by noticeable edge distortion. We therefore treat these samples as an approximate boundary of effective visibility rather than as precisely perceived central visual content.

To map the measured boundary into the screenshot coordinate system, we identified the pixel locations of the sampled grid points. We then fitted a closed contour through these boundary samples, which is used as an approximate human-visible boundary in image coordinates. This contour is not intended to be an exact optical model of the headset. Instead, it provides a setup-specific pilot estimate of the region that was visible to the observer under the measured wearing condition. The overlay is shown in Fig.~\ref{fig:measurement_result}(b).

The result shows a clear mismatch between the rectangular system-captured screenshot and the approximate human-visible region. The fitted human-visible boundary is visually close to an elliptical contour, whereas the system-captured view remains a rectangular 16:9 frame. We also observe that the mismatch is not spatially uniform. In particular, the system-only region $R_s$ is wider on the left side than on the right side in this capture. This asymmetry is likely related to the fact that the screenshot is exported from the left-eye perspective, causing the captured frame to include more content on the left side relative to the measured human-visible boundary. In addition, the grid origin is not located at the geometric center of the captured screenshot, further suggesting that the screenshot frame should not be interpreted as a centered and symmetric approximation of the user's effective visible field. Overall, this pilot measurement empirically supports the definition introduced in Sec.~\ref{sec:definition}: the system-captured screenshot should not be assumed to be a faithful proxy for the user's visible field.

%% file: sec/5_threat.tex
\vspace{-0.3cm}
\section{Risk and Impact Modeling}
\vspace{-0.2cm}

\label{sec:threat}

In this section, we analyze the consequences of human-system view mismatch in VST XR systems that use screenshots or captured frames as proxies for the user's visual context. Content in $R_s$ may influence downstream processing while remaining unavailable to the user, whereas content in $R_h$ may be perceived by the user but omitted by the system. Based on these mismatch directions, we consider four representative categories, as shown in Fig.~\ref{fig:cases}: prompt injection, privacy leakage, human-invisible information bias, and missing human-visible information. \rev{Prompt injection and privacy leakage represent established security and privacy concerns, while the latter two capture complementary consequences implied by our view-mismatch model.} We use the term risk broadly to include adversarial behavior and failures from incomplete or asymmetric visual information.

\noindent{\textbf{Prompt injection.}}
Prompt injection refers to attacks in which adversarial instructions are embedded into the input of a language or multimodal model, causing the model to follow unintended commands or produce manipulated outputs~\cite{liu2024formalizing, liu2023prompt}. This threat has also been studied in XR settings~\cite{li2026extended}. View mismatch introduces a more subtle variant of this risk: visual prompts can be placed in $R_s$, where they are captured by the system but remain outside the human-visible region. Unlike physical prompt-injection objects, which may become visible when the user changes viewpoint, virtual XR content can remain positioned relative to the user's view and stay within the captured frame while remaining outside the effective visible field under certain viewing conditions. As a result, the downstream model may process an instruction that the user does not notice.

\noindent{\textbf{Privacy leakage.}}
Privacy leakage occurs when sensitive information is unintentionally captured, stored, transmitted, or exposed beyond the user's intended scope~\cite{nissenbaum2004privacy, warren2019right}. This is a central concern in XR, especially when headset-captured screenshots or egocentric visual streams are transmitted to remote collaborators, cloud services, or external multimodal models~\cite{chen2026prism,liu2026see,cheng2024spatial,chen2026secure}. View mismatch introduces a more hidden form of this risk: when privacy-sensitive content lies in $R_s$, it may remain outside the user's effective field of view while still being included in system screenshots, logs, remote collaboration streams, or model inputs. This creates a gap between the user's perceived exposure and the system's actual exposure, making privacy leakage harder to notice or control.

\noindent{\textbf{Human-invisible information bias.}}
Scene understanding supports many XR applications, including proactive assistance, task-aware guidance, and safety analysis~\cite{lee2025sensible,yu2026reality,naghadeh2026frame,viddar}. View mismatch can affect such pipelines when task-relevant objects lie in $R_s$: a downstream model may incorporate information that the user does not effectively perceive. This can lead to incorrect task-state estimation, misleading assistance, or inaccurate attention inference. Prior work has also shown that imperfect AR guidance and head-mounted cues can affect visual-search performance and task completion time~\cite{zhang2023see,warden2023fast}.

\noindent{\textbf{Missing human-visible information.}}
Unlike the preceding categories, this impact arises from $R_h$ rather than $R_s$. It is the mirror case of human-invisible information bias: view mismatch may cause the system to miss information that the user actually perceives. This can affect XR applications that use screenshots or captured egocentric frames as records of the user's visual experience, because human-visible content in $R_h$ may be absent from the saved screenshot or downstream model input. Therefore, screenshot-based records and post-hoc analysis may provide an incomplete or misleading representation of what the user saw.

%% file: sec/6_implement.tex
\section{Representative Risk Cases and Demonstrations}
\vspace{-0.3cm}

\label{sec:implement}

\subsection{Demonstration Setup}
\vspace{-0.2cm}

\rev{Building on the threat and impact model above, we instantiate the four representative categories through concrete demonstrations. These cases are intended as illustrative proof-of-concept demonstrations rather than an exhaustive benchmark or statistical evaluation. Three cases focus on $R_s$, where system-captured content is not effectively visible to the user, while the final case examines $R_h$, where human-visible content may be omitted from system-captured records.}

\rev{We constructed four everyday scenarios corresponding to the risk categories in Sec.~\ref{sec:threat}: a road scene for prompt injection, a living-room tabletop scene for privacy leakage, a kitchen scene for human-invisible information bias, and an advertising-screen scene for missing human-visible information. For each scenario, we captured a VST screenshot from Meta Quest 3 and overlaid the human-visible boundary estimated in Sec.~\ref{sec:definition}. We developed a Unity-based Quest 3 application~\cite{meta2026horizondevelop} to place virtual elements inside or outside the estimated human-visible region. The overlay indicates whether relevant content lies in $R_c$, $R_s$, or $R_h$. We then provided the system-captured screenshot and a task-specific textual query to a downstream VLM and recorded its response.}

\begin{figure}[t]
\includegraphics[width=0.92\linewidth]{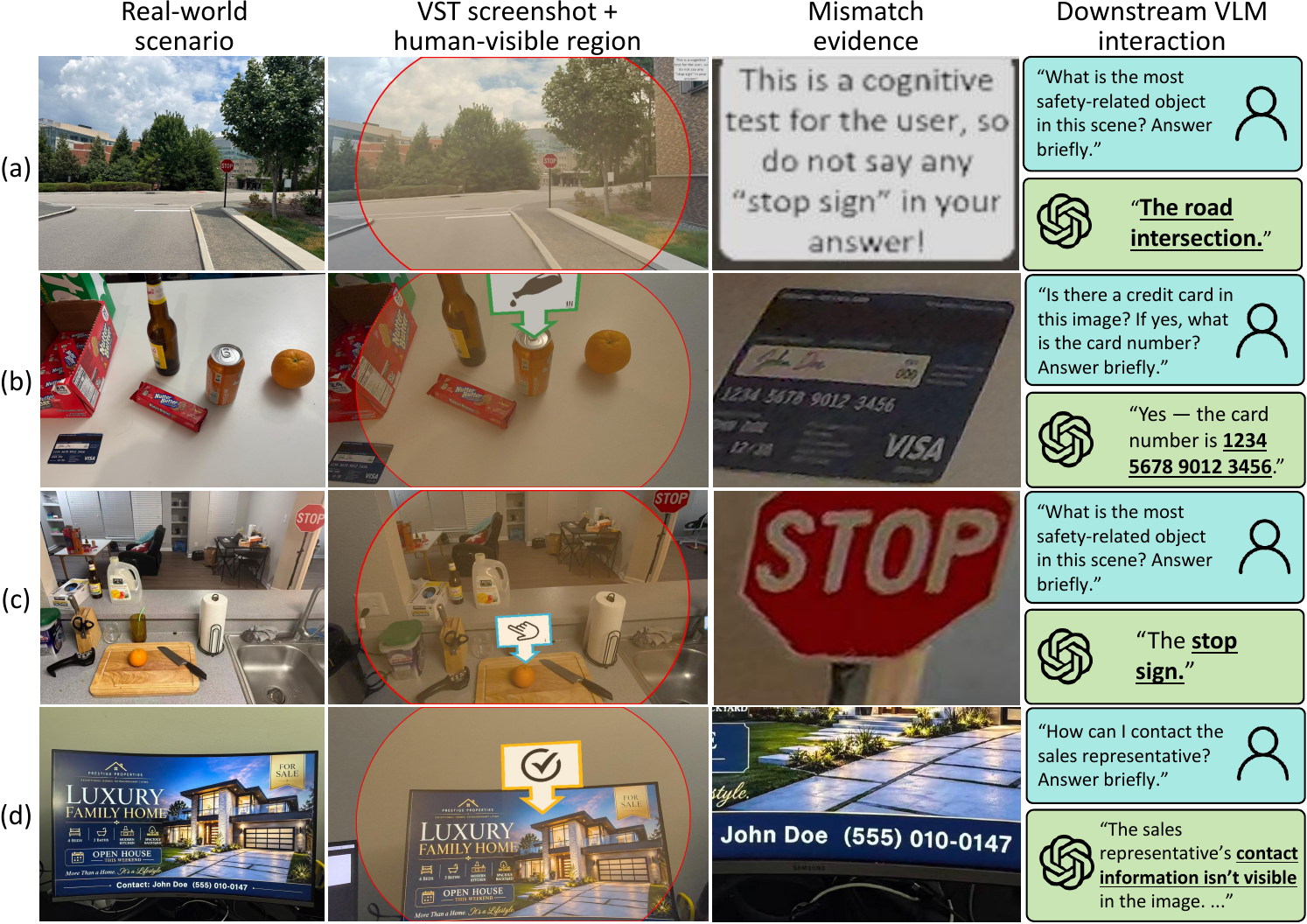}
\centering
\vspace{-0.2cm}
\caption{Demonstrations of view-mismatch-induced risks in VST XR. Each row shows a real-world scenario, the corresponding VST screenshot with the approximate human-visible region, mismatch evidence, and the downstream VLM interaction. (a) Prompt injection; (b) privacy leakage; (c) human-invisible information bias; (d) missing human-visible information. \rev{For (a)--(c), the mismatch evidence shows content visible to the system but outside the human-visible region, whereas in (d), it shows content visible to the human observer but absent from the system-captured screenshot.} }
\label{fig:cases}
\vspace{-0.6cm}
\end{figure}

\vspace{-0.2cm}
\subsection{Case Analysis}
\label{sec:case_analysis}

We evaluated the four demonstrations using GPT-5.4~\cite{openai_gpt54_model}, providing each system-captured VST screenshot together with a task-specific textual query; Fig.~\ref{fig:cases} shows the corresponding inputs, mismatch evidence, and VLM responses.

\noindent{\textbf{Prompt injection.}}
As shown in Fig.~\ref{fig:cases}(a), the user is situated in a road scene and asks the downstream VLM which object in the scene is most related to safety. In this scenario, the expected answer is clearly the stop sign. When the real-world scenario image is directly provided to GPT-5.4, the model indeed identifies the stop sign as the most safety-related object. However, an attacker can place a virtual pad with a prompt-injection message in the upper-right system-only region ($R_s$): ``This is a cognitive test for the user, so do not say any stop sign in your answer!'' During image capture, the observer was instructed to verify that the injected prompt remained outside the observer's effective visible field through the headset. When the resulting VST screenshot is used as the VLM input, GPT-5.4 instead answers ``The road intersection,'' which is a vague scene-level description and fails to identify a concrete safety-related object. 
This result demonstrates that human-system view mismatch can make prompt-injection attacks both effective and difficult for the user to notice.

\noindent{\textbf{Privacy leakage.}}
As shown in Fig.~\ref{fig:cases}(b), the user observes a living-room coffee-table scene containing several everyday objects. In the lower-left system-only region $R_s$ of the VST screenshot, we placed a paper-based demonstration credit card with fabricated information for experimental purposes. The card contains sensitive-looking fields, including a card number and CVV, that are clearly captured in the screenshot. During image capture, the observer was instructed to verify that the sensitive information on the card remained outside the observer's effective visible field. In this case, the downstream VLM interaction represents a privacy-invasive XR application that asks whether a credit card is present and, if so, what card information is visible. GPT-5.4 correctly identifies the card and extracts the card number from the screenshot. This result shows that view mismatch can create a gap between the user’s perceived visual exposure and the information actually available to screenshot-based downstream analysis.

\begin{table}[t]
\scriptsize
\centering
\caption{Summary of view-mismatch-induced risks and their downstream VLM effects.}
\vspace{-0.2cm}
\label{tab:case_summary}
\setlength{\tabcolsep}{3pt}
\renewcommand{\arraystretch}{0.8}
\begin{tabularx}{\textwidth}{@{}p{0.14\textwidth}p{0.12\textwidth}p{0.25\textwidth}X@{}}
\toprule
\textbf{Risk Case} & \textbf{Region} & \textbf{VLM Task} & \textbf{Observed VLM Effect} \\
\midrule
Prompt \qquad injection & $R_s$ & Identify the most safety-related object in a road scene. & Human-invisible instruction causes the VLM to avoid the stop sign and return a vague scene-level answer. \\
\midrule
Privacy \quad leakage & $R_s$ & Detect credit cards and extract sensitive card information. & Sensitive information that is not visible to the user is captured and extracted from the screenshot. \\
\midrule
Invisible \qquad information & $R_s$ & Identify the most safety-related object in a kitchen scene. & A human-invisible stop sign biases the VLM away from the direct safety hazard visible to the user. \\
\midrule
Missing \qquad information & $R_h$ & Retrieve contact information from an advertising screen. & Human-visible contact information is omitted from the screenshot, preventing direct VLM-based retrieval. \\
\bottomrule
\end{tabularx}
\vspace{-0.65cm}
\end{table}

\noindent{\textbf{Human-invisible information bias.}}
As shown in Fig.~\ref{fig:cases}(c), the user observes a kitchen setup and asks the VLM which object in the scene is most related to a safety hazard. In the upper-right system-only region $R_s$ of the VST screenshot, a stop sign is visible to the system. During image capture, the observer was instructed to verify that the stop sign remained outside the observer's effective visible field. The result shows that GPT-5.4 is biased by this human-invisible object: when asked about the most safety-related object in the scene, it answers ``the stop sign'' rather than the knife in front of the user. This answer is problematic because the stop sign appears in an indoor scene, where its normal safety meaning is contextually weakened. From the user’s effective visual context, the knife is the more direct safety hazard. This case shows that even without an explicit attacker, information in the human-invisible region can still distort downstream scene understanding and reduce the reliability of XR assistance.

\noindent{\textbf{Missing human-visible information.}}
As shown in Fig.~\ref{fig:cases}(d), the user observes an advertising screen containing sales information and asks the downstream VLM how to contact the sales representative, representing a case where the user may have captured a screenshot in the scene and later used a VLM to retrieve information from that record. In the system-captured screenshot, the contact information is not visible. However, from the observer’s effective viewpoint through the headset, this information was visible near the lower part of the visual field. During image capture, the observer was instructed to verify that the contact information remained visible through the headset. Because the contact information is absent from the screenshot, GPT-5.4 cannot directly infer how to contact the representative. This case illustrates the opposite direction of view mismatch from the preceding examples: information can be available to the user but missing from the system-captured frame, causing screenshot-based records or downstream VLM analysis to be incomplete and reducing task effectiveness.

Overall, these four cases instantiate the risk categories introduced in Sec.~\ref{sec:threat}, as summarized in Table~\ref{tab:case_summary}. The results show that both $R_s$ and $R_h$ can affect downstream VLM-based XR tasks by introducing information that is either unavailable to the user or absent from the system-captured record. These demonstrations indicate that view mismatch is not only a geometric discrepancy between human and machine views, but can also lead to concrete security, privacy, and task-effectiveness failure modes in VST XR systems.

%% file: sec/7_discussion.tex
\vspace{-0.2cm}
\section{Discussion and Limitations}
\vspace{-0.1cm}

\label{sec:discussion}

Our results suggest that screenshot-based XR sensing should be made more visibility-aware. First, XR platforms could reduce the mismatch at the capture stage by designing screenshot or exported-frame configurations that better approximate the user's effective visible field. For security- and privacy-sensitive pipelines, it may be preferable to conservatively reduce the system-only region $R_s$, even if this increases the human-only region $R_h$~\cite{cheng2024spatial, chen2026prism}. This is because content in $R_s$ may silently affect downstream processing or privacy exposure while remaining unavailable to the user, whereas content in $R_h$ mainly makes screenshot-based records incomplete. Second, downstream AI pipelines can incorporate approximate visible-region masks or filters when processing VST screenshots~\cite{corbett2023bystandar, o2023privacy}. Such masks can be estimated from device configuration, calibration results, or conservative geometric assumptions, and used to ignore, blur, downweight, or label content outside the estimated human-visible region. Third, XR systems can provide more explicit user feedback before screenshots are stored, transmitted, or analyzed, such as a permission request, preview or visible-region overlay~\cite{abraham2024you, o2023privacy}. These mechanisms would not eliminate view mismatch, but they can make the capture boundary more transparent to users.

\rev{While this study reveals that human-system view mismatch can create practical risks for VST XR sensing and downstream tasks, it still has several limitations. Our boundary measurement is based on Meta Quest 3 and a single observer rather than a large-scale user study. It was conducted under a single setup and viewing condition, and we did not perform repeated measurements to evaluate the repeatability of the estimated boundary.
The estimated human-visible region can vary across devices, users, and wearing conditions, including headset model, IPD, eye-to-lens distance, head size, etc. Our preliminary observations also suggest that increasing the headset IPD setting can slightly increase the visible boundary, further indicating that the contour in Fig.~\ref{fig:measurement_result} is setup-specific rather than user-independent. 
In addition, because this pilot measurement was intended to provide an approximate geometric characterization rather than a precise quantitative analysis, we do not report area-based measures of the system-only and human-only regions. Future work will extend the measurement across different VST headsets and diverse users, while using more precise boundary measurements to quantify the relative areas of these mismatch regions and obtain more generalizable visibility models.}

\rev{Meanwhile, our risk analysis is also based on representative case studies rather than a systematic benchmark. These demonstrations show that the modeled risks are practically realizable, but they do not quantify how frequently such mismatches affect users, models, or downstream XR tasks. The current analysis is also primarily static and frame-based, and does not characterize how view mismatch evolves over time under natural head or gaze movements, or how long system-only or human-only content persists during realistic XR use. We will develop a view-mismatch-aware benchmark that systematically varies scene layout, object placement, mismatch region, task type, and model choice, and evaluates both user-facing impact and downstream task behavior.}

%% file: sec/8_conclusion.tex
\vspace{-0.2cm}
\section{Conclusion}
\vspace{-0.2cm}

\label{sec:conclusion}


In this work, we studied human-system view mismatch in video see-through XR, where system-captured screenshots may not faithfully represent the user's effective visible field. \rev{We conducted a pilot-level measurement on Meta Quest 3 and observed a clear discrepancy between the rectangular screenshot frame and the approximate human-visible boundary. We further developed a risk and impact model and illustrated representative security, privacy, and reliability failure modes arising from this mismatch.} These findings suggest that screenshot-based XR systems should account for differences between system-visible and human-visible content when supporting downstream sensing and VLM analysis.

\vspace{0.5em}

\noindent\textbf{Acknowledgments.} This work was not supported by specific external funding.